# Estimating the number of serial killers that were never caught

M.V. Simkin and V.P. Roychowdhury
Department of Electrical and Computer Engineering, University of California, Los Angeles, CA 90095-1594

**Abstract.** Many serial killers commit tens of murders. At the same time inter-murder intervals can be decades long. This suggests that some serial killers can die of an accident or a disease, having been never caught. We use the distribution of the killers by the number of murders, the distribution of the length of inter-murder intervals and USA life tables to estimate the number of the uncaught killers. The result is that in 20th century there were about seven of such killers. The most prolific of them likely committed over sixty murders.

In a recent paper (Yaksic et al , 2021) we analyzed the data on 1,012 US serial killers who were active in 20th century. We found that the distribution of the killers by the numbers of murders they committed follows a power law. We found that the distribution of inter-murder intervals also follows a power law. A power law is a highly skewed distribution so select serial killers commit tens of murders and some inter-murder intervals are decades long. These results imply that some serial killers could have just died of an accident or a disease and had been never caught. Interestingly one can use statistical data to estimate the number of such killers.

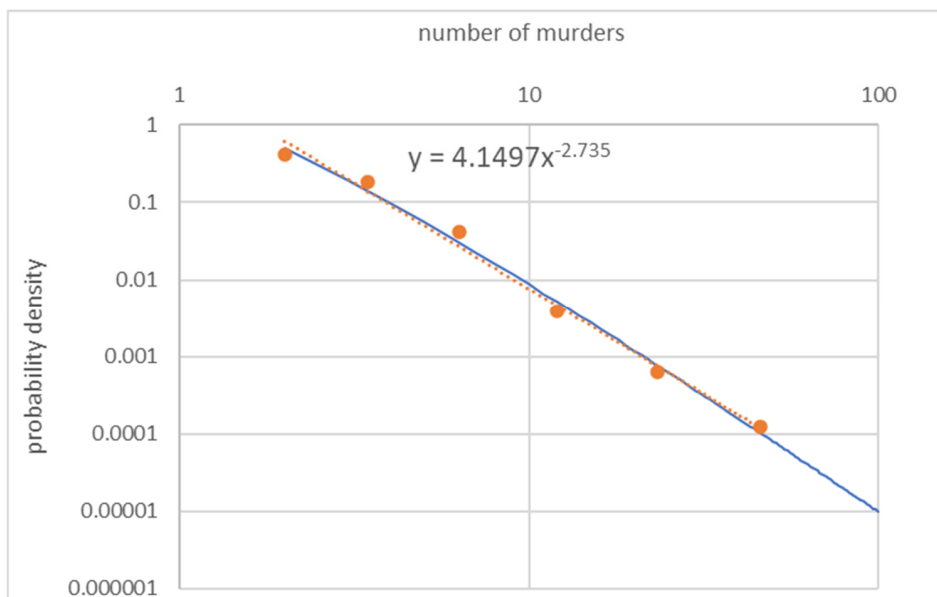

*Figure 1.* The distribution of 1,012 US serial killers by the number of murders they committed. Solid circles – the actual data. Dotted line - a least-square fit using a power law. Solid line – maximum likelihood fit using Eq.(3).

Figure 1 shows the distribution of 1,012 US serial killers by the number of murders they committed. Can we come up with a model which would reproduce such a distribution? Actually, we already did something similar in the study of the victory scores of fighter-pilot aces (Simkin & Roychowdhury, 2008). Let us suppose that with probability $r$ the killer murders without a blunder. With the probability $1 - r$ he blunders and gets caught. The probability to commit exactly $n$ murders before getting caught is

$$p(n) = (1 - r) \times r^{n-1}$$

This gives an exponential distribution of the killers by the number of murders. However, this is only true if $r$ is the same for every killer.

If we, for example, assume that $r$ is uniformly distributed between 0 and 1 we get

$$p(n) = \int_0^1 dr(1-r) \times r^{n-1} = \frac{1}{n \times (n+1)} \approx n^{-2}$$

This is already a power law just as in Figure 1, only the exponent is different. If we now suppose that the distribution of $r$ is not uniform but depleted in the region of the most skillful killers

$$p(r) = (1+\alpha) \times (1-r)^\alpha \tag{1}$$

We get

$$p(n) = (1+\alpha) \times \int_0^1 dr(1-r)^{1+\alpha} \times r^{n-1} = (1+\alpha) \times \mathrm{B}(n, 2+\alpha) \tag{2}$$

Here B is Euler's Beta function. Using the large $n$ asymptotic of Beta function we get

$$p(n) \approx (1+\alpha) \times \Gamma(2+\alpha) \times n^{-2-\alpha}$$

Here $\Gamma$ is Euler's Gamma function. If we take $\alpha = 0.735$ we get the same slope as the power-law fit in Figure 1[1]. One may argue that it is not reasonable to use the large-$n$ asymptotic in the region of small $n$. Indeed if we discard two left-most data points in Figure 1 and re-do the power-law fit using four remaining points we get $\alpha = 0.918$. A simple power-law fit is easy to do since it is a built-in feature in Excel. So it was a reasonable thing to start with it. However, it is more rigorous to use Eq.(2) to do the fitting. The least square fit is not the best thing to do since the results depend on the way we bin the data.

A more scientific approach is the maximum likelihood estimate. The method is rather straightforward except for one thing. The distribution we work with includes only successful serial killers, that is those who committed at least two murders. Eq.(2) also includes the would-be serial killers, those arrested after their very first murder. So we should substitute for Eq.(2) the conditional probability with the condition that the number of murders is more than 1:

$$p(n|n>1) = p(n)/(1-p(1)) = (1+\alpha) \times \mathrm{B}(n, 2+\alpha)/(1-(1+\alpha) \times B(1, 2+\alpha)) \tag{3}$$

Numerical maximization of the likelihood with regard to $\alpha$ gives $\alpha = 0.945$. Probability distribution computed using Eq.(3) is shown in Fig.(1) as solid line.

We have mentioned that earlier we tackled a similar problem regarding fighter pilot aces: obtaining the underlining distribution of the success rates which results in the observed distribution of victory scores (Simkin & Roychowdhury, 2008). There were two categories of fighter pilots, however. Those who were finally shot down and killed or taken prisoner, and those who survived the war undefeated. We could guess that similarly there should be serial killers that were never caught. Albeit, unlike the former, the latter do not hurry to announce themselves.

In the aces study (Simkin & Roychowdhury, 2008) it was straightforward to estimate the defeat rates of finally defeated pilots. If the pilot shot down $n$ opponent aircraft before being shot down himself, he had 1 defeat out of $n+1$ decisive engagements. So the defeat rate is just $1/(n+1)$. It is trickier to do this for undefeated pilots. One, however, can use the distribution of defeat rates of defeated pilots as a prior, and make a Bayesian estimate of the defeat rate of an undefeated pilot using the fact he had not lost in *n* decisive engagements. Of course, one can use the same technique to make a better estimate of the defeat rate of defeated pilots then the aforementioned $1/(n + 1)$.

---

[1] The reader may notice that this is a different number from what was in Figure 1 of (Yaksic et al, 2021). This is because in that paper we associated the estimated probability density with the upper boundaries of the bins. In present paper we associated it with the geometric mean of bin boundaries (which is the middle of the bin on logarithmic scale).

A similar technique was used before to estimate the dropped call rates of cellular phone service providers in the case when in an actual test it was found to be zero (Simkin & Olness, 2002). And much earlier it was used to make improved estimates of batting averages of baseball players and many other problems (Efron and Morris, 1977). Here we have a similar but more challenging and more intriguing problem: to find the actual number of hidden serial killers.

The most straightforward way is to do a numerical simulation. We will assume that the success rates of all serial killers come from the distribution given by Eq. (1) with the exponent determined by the maximum likelihood estimate $\alpha = 0.918$. We will use the actual inter-murder intervals in the simulation by drawing one of them at random at each step. The killers in the simulation will die according to 1950 USA life tables (Bell and Miller, 2005). Since 1950 is the middle of the century we are interested in it seems a reasonable lifetable to use.

We will also obviously need the age at first murder to determine the age of the killer during the simulation. We did not report the distribution of killers by the age at first murder in the previous paper (Yaksic et al, 2021). However, this statistic is available in the same database (Yaksic 2015). We looked at it now. It is approximately lognormal. However, we will not use the functional form of the distribution in the simulation. We will just draw the age at random out of all actual ages at first murder.

The algorithm is as follows. For each simulated killer we first draw the success rate *r* from the distribution given by Eq.(3). On the technical side we compute the cumulative density function, equate it to a uniformly distributed random number and invert the equation to find *r*. So, the actual formula that we use is $r = 1 - (1 - rand)^{1/(\alpha+1)}$ where *rand* is a uniformly distributed on the interval [0,1) random number. Next assign the age $x_1$ to the killer by drawing at random from the actual ages at first murder. Next, we generate another random number *rand* and if $rand > r$ the killer is caught after the first murder. We record that the killer committed 1 murder and that he was caught. And proceed to simulate the next killer. In the case $rand < r$ the killer was not caught after first murder. We select the first inter-murder interval by drawing a random interval out of the array of actual inter-murder intervals. We compute the new age $x_2$ of the killer by adding the selected inter-murder interval to $x_1$. Then we use the column $l_x$ from 1950 USA life tables. The ratio $l_{x_2}/l_{x_1}$ is the conditional probability that the person will be alive at age $x_2$ provided he was alive at age $x_1$. We use another random number to decide whether the killer will be alive at the time he is to commit his second murder. If he is dead, we record that the killer committed one murder and that he was uncaught. If he is alive, we select next random inter-murder interval and repeat all already described steps. We do it until the killer either gets caught or dies.

We did this Monte-Carlo simulation for a sample of 1,000,000 killers. Out of those 659,684 got caught after first murder. 539 died after the first murder. 337,729 killers succeeded to commit 2 or more murders and thus became serial killers but finally were caught. 2,048 committed 2 or more murders and were not caught. The distribution of the simulated caught and uncaught serial killers is shown in Figure 2. One can see that the distribution of the uncaught killers has a quite high density on the large number of murders end.

So how many serial killers went uncaught in 20[th] century? The ratio of uncaught to caught killers in the simulated sample was $2,048/337,729 = 0.006064$. We would expect the same ratio in the real sample. In the beginning of the article, we mentioned 1,012 20[th] century US serial killers. It was also the sample studied in (Yaksic et al , 2021). However, it was a reduced sample for it included only those killers for which we knew the exact date of each of their murders (it is important when you study inter-murder intervals). The complete sample, however, includes 1,172 serial killers. So we expect $0.006064 \times 1{,}172 = 7.107$ uncaught serial killers. This is not a whole number, but what it actually means is that the actual number comes from a Poisson distribution with the aforementioned mean. This distribution is shown in Figure 3. We also estimate that with 90% probability the most prolific of the uncaught killers committed at least 20 murders and with 50% probability he committed 60 or more murders.

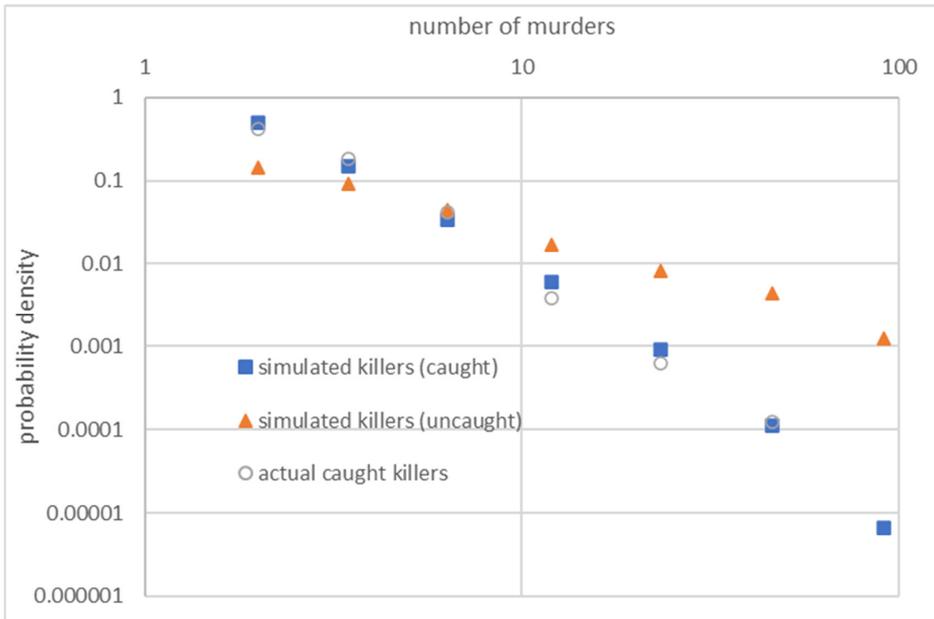

Figure 2. Distribution of simulated caught and uncaught killers together with the distribution of the actual caught killers.

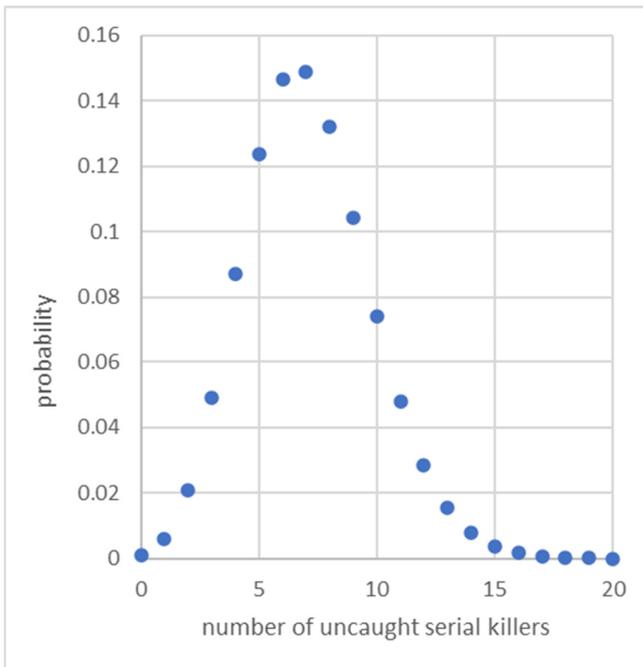

Figure 3. Estimated number of 20[th] century uncaught serial killers.

Note also that our results imply that in the 20$^{th}$ century there were about $659,684/337,729 \times 1,172 = 2,289$ would be serial killers. Those who would become such had they not been arrested after first murder.

The study has an obvious drawback: the use of USA life tables. A person can be technically alive but in such poor health that he is not able to commit a murder. Ideally one would use not life tables but something like "active life tables" would they be available. Note, however, that active life span is always smaller than just life span. So the fraction of the uncaught killer would be only bigger. So our estimate is on a lower side.